\documentclass{river-journal}
\usepackage{rivps}

\usepackage{graphicx}
\usepackage{graphics}
\usepackage{subfigure}
\usepackage{cite}

\par

\begin{document}
\begin{opening}
\title{Hidden Risks: The Centralization of NFT Metadata and What It Means for the Market}
\author{Hamza Salem}
\institute{Innopolis University \texttt{h.salem@innopolis.university}}
\end{opening}

\runningtitle{Sample Document}
\runningauthor{H.Salem}

\subsection*{Abstract}
The rapid expansion of the non-fungible token (NFT) market has catalyzed new opportunities for artists, collectors, and investors, yet it has also unveiled critical challenges related to the storage and distribution of associated metadata. This paper examines the current landscape of NFT metadata storage, revealing a significant reliance on centralized platforms, which poses risks to the integrity, security, and decentralization of these digital assets. Through a detailed analysis of top-selling NFTs on the OpenSea marketplace, it was found that a substantial portion of metadata is hosted on centralized servers, making them susceptible to censorship, data breaches, and administrative alterations. Conversely, decentralized storage solutions, particularly the InterPlanetary File System (IPFS), were identified as a more secure and resilient alternative, offering enhanced transparency, resistance to tampering, and greater control for creators and collectors. This study advocates for the widespread adoption of decentralized storage architectures, incorporating digital signatures to verify ownership, as a means to preserve the value and trustworthiness of NFTs in an increasingly digital world. The findings underscore the necessity for NFT platforms to prioritize decentralized methodologies to ensure the long-term sustainability and integrity of the NFT ecosystem.


\section{Introduction}
The non-fungible token (NFT) market has experienced exponential growth in recent years, offering new opportunities for artists, collectors, and investors. NFTs, which are unique digital assets verified using blockchain technology, have redefined the concept of ownership in the digital space. However, alongside the burgeoning market, critical issues have emerged, particularly concerning the storage and distribution of metadata associated with these digital assets.

Metadata is crucial to the value and functionality of NFTs as it often contains essential information about the asset, including its origin, ownership, and visual or auditory data. Despite the decentralized nature of blockchain technology, which underpins NFTs, the metadata itself is frequently hosted on centralized platforms. This reliance on centralized servers for metadata storage introduces several vulnerabilities that undermine the decentralization ethos of NFTs. These include security risks, such as data breaches and hacking, as well as potential censorship and loss of control for artists and collectors.

This paper seeks to address these challenges by examining the current state of metadata storage in the NFT ecosystem. Through an analysis of data collected from top-selling NFTs on the OpenSea marketplace, this study evaluates the extent to which metadata is hosted on centralized platforms versus decentralized solutions like the InterPlanetary File System (IPFS). The findings highlight the urgent need for NFT platforms to adopt decentralized storage architectures to enhance the security, transparency, and resilience of NFTs.

Moreover, this paper advocates for the development of a new storage architecture that incorporates digital signatures to verify ownership, ensuring that only verified owners can access or modify associated metadata. By doing so, the NFT ecosystem can preserve the integrity of these digital assets, thereby safeguarding their value and trustworthiness in the marketplace.

\section{Literature Review}

The rapid growth of the non-fungible token (NFT) market has attracted considerable academic and industry attention, particularly concerning the implications of decentralized and centralized metadata storage solutions. The literature in this domain primarily focuses on the technical, economic, and security aspects of NFTs, with an increasing emphasis on the challenges and risks associated with metadata storage.

\subsection{NFTs and Blockchain Technology}

NFTs are digital assets authenticated by blockchain technology, which ensures the uniqueness and ownership of these assets. \cite{chen2021nft} discuss the fundamental principles of NFTs and their reliance on blockchain for verification. The decentralized nature of blockchain is a cornerstone of NFTs, ensuring that ownership records are immutable and transparent.

However, while the blockchain itself is decentralized, \cite{regner2019nft} highlight that the metadata associated with NFTs is often stored off-chain, frequently on centralized servers. This practice contradicts the decentralized promise of blockchain, as it exposes NFTs to risks associated with centralization, including data breaches, loss of control, and potential censorship.

\subsection{Challenges in Metadata Storage}

The issue of centralized metadata storage has been critically examined in recent studies. \cite{kraken2021nft} emphasize that centralized platforms, such as Amazon Web Services (AWS), are commonly used to host NFT metadata. This reliance on centralized services poses significant risks, as these platforms are vulnerable to hacking, data breaches, and unilateral administrative changes.

On the other hand, \cite{benet2014ipfs} propose the InterPlanetary File System (IPFS) as a decentralized alternative for metadata storage. IPFS is a peer-to-peer network that allows data to be stored across a distributed network of nodes, enhancing security and resilience. The use of IPFS ensures that metadata is not dependent on any single server, thereby reducing the risks associated with centralization.

\subsection{Security and Integrity of NFTs}

The integrity and security of NFTs are paramount to maintaining their value and trustworthiness. \cite{gupta2021nftsecurity} argue that decentralized storage solutions, such as IPFS, not only enhance security but also offer greater transparency and control for artists and collectors. Decentralized storage makes it difficult for bad actors to tamper with or corrupt metadata, thereby preserving the authenticity of NFTs.

Moreover, the literature suggests that incorporating digital signatures as part of the storage architecture can further enhance the security of NFTs. \cite{nakamoto2008bitcoin} originally introduced the concept of digital signatures in the context of blockchain, and this approach can be adapted to ensure that only verified owners can access or modify the metadata associated with their NFTs.

\subsection{Emerging Trends in NFT Metadata Storage}

Recent research has also explored alternative decentralized storage solutions beyond IPFS. \cite{wood2014ethereum} discuss the potential of using blockchain itself for metadata storage, although this approach is still in its nascent stages due to concerns about scalability and cost.

The findings of \cite{voshmgir2017blockchain} suggest that while IPFS is currently the most popular decentralized solution, the NFT ecosystem is gradually exploring other decentralized storage methods that may offer even greater security and efficiency.

\subsection{Conclusion}

The literature underscores the critical need for decentralized metadata storage in the NFT market to uphold the principles of security, transparency, and decentralization. While IPFS presents a viable solution, ongoing research is essential to explore alternative methods and develop more robust storage architectures that can safeguard the integrity of NFTs.

\section{Methodology}

This study aims to investigate the decentralization of metadata storage in the non-fungible token (NFT) ecosystem, focusing on top-selling NFTs from the OpenSea marketplace. The methodology is designed to assess the extent to which NFT metadata is hosted on centralized versus decentralized platforms, thereby evaluating the associated risks and potential vulnerabilities. The study follows a systematic approach that includes data collection, script construction for metadata analysis, and statistical evaluation.

\subsection{Data Collection}

The first step in the research process involved selecting a representative sample of NFT collections from OpenSea, one of the largest and most active NFT marketplaces. The focus was on top-selling NFT collections, as these are likely to have a significant impact on the broader market. To ensure a comprehensive analysis, we selected a diverse range of collections that vary in size, popularity, and market capitalization.

A web scraping tool, specifically the Python library \texttt{BeautifulSoup}, was employed to extract data from the OpenSea platform. The script was designed to capture key information for each NFT within the selected collections, including the token ID, contract address, metadata URI, and the storage platform where the metadata is hosted. The following is an outline of the data collection process:

\begin{itemize}
    \item \textbf{Collection Selection:} Top-selling NFT collections were identified based on sales volume and popularity on OpenSea.
    \item \textbf{Data Extraction:} For each NFT within the selected collections, the token ID and contract address were extracted.
    \item \textbf{Metadata Retrieval:} Using the extracted contract address, the metadata URI was retrieved from the Ethereum blockchain via \texttt{etherscan.io}.
    \item \textbf{Storage Identification:} The metadata URI was analyzed to determine whether the metadata is stored on-chain, off-chain, or using decentralized solutions such as IPFS.
\end{itemize}

\subsection{Script Construction}

The metadata analysis script was developed in Python, leveraging the \texttt{web3.py} library to interact with the Ethereum blockchain and the \texttt{requests} library for HTTP requests. The script was designed to automate the process of determining the storage location of NFT metadata. The steps involved in the script construction are outlined below:

\begin{enumerate}
    \item \textbf{Blockchain Interaction:} The script interacts with the Ethereum blockchain using the \texttt{web3.py} library to retrieve metadata URIs from the NFT contract addresses.
    \item \textbf{Metadata URI Parsing:} The retrieved metadata URIs are parsed to identify whether they point to a decentralized network (e.g., IPFS) or a centralized server (e.g., AWS, Google Cloud).
    \item \textbf{Storage Categorization:} Based on the analysis of the metadata URI, the script categorizes each NFT as being hosted on-chain, off-chain (centralized), or off-chain (decentralized, e.g., IPFS).
    \item \textbf{Output Generation:} The script generates a summary report, detailing the proportion of NFTs within the sample that are stored on-chain, off-chain (centralized), or off-chain (decentralized).
\end{enumerate}

% \subsection{Survey Design}

% In addition to the automated script, a manual survey was conducted to validate the script's results and to gather qualitative insights from NFT platforms and artists. This survey was designed to understand the rationale behind the choice of storage solutions for NFT metadata. The survey included the following key components:

% \begin{itemize}
%     \item \textbf{Respondents:} The survey targeted NFT creators, platform administrators, and collectors who are directly involved in the storage and management of NFT metadata.
%     \item \textbf{Questions:} The survey included questions about the respondents' preferences for metadata storage (on-chain vs. off-chain), their awareness of the risks associated with centralized storage, and their experiences with decentralized storage solutions like IPFS.
%     \item \textbf{Analysis:} The responses were analyzed to identify trends and preferences in metadata storage, which were then compared with the quantitative findings from the automated script.
% \end{itemize}

\subsection{Statistical Analysis}

The data collected through the automated script  was subjected to statistical analysis to determine the proportion of NFTs that are stored in a centralized versus decentralized manner. The results were categorized as follows:

\begin{itemize}
    \item \textbf{Decentralized (IPFS):} NFTs with metadata hosted on decentralized platforms like IPFS.
    \item \textbf{Centralized:} NFTs with metadata hosted on centralized servers such as AWS or Google Cloud.
    \item \textbf{On-Chain:} NFTs with metadata fully stored on the blockchain.
    \item \textbf{Uncertain:} NFTs where the storage method could not be clearly identified.
\end{itemize}

The findings were then compared with the survey responses to provide a comprehensive view of the current state of metadata storage in the NFT ecosystem. This comparison highlights the gap between preferred practices and the actual implementation of storage solutions in the market.

\section{Results}

The results of this study provide a comprehensive overview of the current state of metadata storage in the non-fungible token (NFT) ecosystem. The analysis was conducted on a sample of top-selling NFTs from the OpenSea marketplace, focusing on the categorization of metadata storage across different platforms. The findings are summarized below.

\subsection{Metadata Storage Distribution}

Based on the analysis conducted using the automated script and manual validation, the metadata associated with NFTs was categorized into four distinct storage types: decentralized (IPFS), centralized, on-chain, and uncertain. The distribution of these categories is as follows:

\begin{itemize}
    \item \textbf{Decentralized (IPFS):} Approximately 38.84\% of the NFTs in the sample had their metadata stored on the InterPlanetary File System (IPFS). This indicates a significant adoption of decentralized storage solutions, which align with the ethos of blockchain technology by providing enhanced security and resistance to censorship.
    
    \item \textbf{Centralized:} A notable 31.68\% of the NFTs had their metadata hosted on centralized platforms, such as Amazon Web Services (AWS) or Google Cloud. This finding highlights a considerable reliance on centralized storage, which poses risks related to data breaches, censorship, and loss of control for NFT creators and owners.
    
    \item \textbf{On-Chain:} Only 25.62\% of the NFTs had their metadata fully stored on-chain. On-chain storage is the most secure method as it ensures that the metadata is immutable and permanently linked to the blockchain. However, the low percentage suggests that on-chain storage is not widely used, likely due to scalability and cost issues.
    
    \item \textbf{Uncertain:} For 3.86\% of the NFTs, the storage method could not be clearly identified. This category includes cases where the metadata URI did not provide enough information to determine the exact storage location or where the data was stored in a non-standard format.
\end{itemize}

\begin{figure}
\centerline{\includegraphics[width=3in]{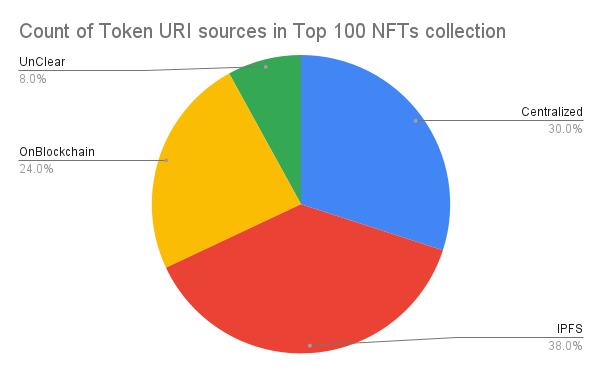}}
\caption[]{Pie chart Show the percentage of the top 100 NFTs collection by volume}%
\label{bridge}
\end{figure}
\section{Conclusion}

This study reveals that while there is a growing adoption of decentralized storage solutions like IPFS in the NFT ecosystem, a significant portion of NFTs still rely on centralized platforms. This reliance poses potential risks to the security, longevity, and control of NFT metadata. The findings underscore the importance of further embracing decentralized approaches to fully realize the benefits of blockchain technology. Addressing the current barriers to on-chain and decentralized storage will be crucial for enhancing the resilience and trustworthiness of NFTs in the future.

\end{document}